\documentclass[preprint,3p,12pt]{elsarticle}

\journal{New Astronomy Review}

\newcommand{\microns}{\ensuremath{\mu\mbox{m}}}

\begin{document}

\begin{frontmatter}



\title{Interferometric science results on young stellar objects}


\author{F.~Malbet}

\address{Laboratoire d'Astrophysique de Grenoble,\\ 
Universit\'e J.~Fourier/CNRS, BP 53, F-38041 Grenoble cedex 9, France}
\ead{Fabien.Malbet@obs.ujf-grenoble.fr}

\begin{abstract}
  Long-baseline interferometry at infrared wavelengths allows the
  innermost regions around young stars to be observed.  These
  observations directly probe the location of the dust and gas in the
  disks.  The characteristic sizes of these regions found are larger
  than previously thought. These results have motivated in part a new
  class of models of the inner disk structure, but the precise
  understanding of the origin of these low visibilities is still in
  debate. Mid-infrared observations probe disk emission over a larger
  range of scales revealing mineralogy gradients in the disk. Recent
  spectrally resolved observations allow the dust and gas to be
  studied separately showing that the Brackett gamma emission can find
  its origin either in a wind or in a magnetosphere and that there is
  probably no correlation between the location of the Brackett gamma
  emission and accretion. In a certain number of cases, the very high
  spatial resolution reveals very close companions and can determine
  their masses. Overall, these results provide essential information
  on the structure and the physical properties of close regions
  surrounding young stars especially where planet
  formation is suspected to occur.
\end{abstract}


\end{frontmatter}



\section{Introduction}

Many physical phenomena occur in the inner regions of the disks which
surround young stars. The matter which eventually falls onto the
stellar surface works its way through an accreting circumstellar disk
which is subject to turbulence, convection, external and internal
irradiation. The disks, which are rotating in a quasi-Keplerian motion,
are probably the birth location of future planetary systems.  Strong
outflows, winds and even jets find their origin in the innermost
regions of many young stellar systems. The mechanisms of these
ejection processes are not well understood but they are probably
connected to accretion. Most stars are born in
multiple systems which can be very tightly bound and therefore have a strong
impact on the physics of the disk inner regions.

The details of all these physical processes are not well understood yet
because of lack of data to constrain them. The range
of physical parameters which define best the inner disk regions in
young stellar objects are:
\begin{itemize}
\item radius ranging from 0.1\,AU to 10\,AU
\item temperature ranging from 150\,K to 4000\,K
\item velocity ranging from 10\,km/s to a few 100\,km/s
\end{itemize}

The instrumental requirements to investigate the physical conditions
in such regions are therefore driven by the spectral coverage which
must encompass the near and mid infrared from 1 to $20\,\microns$.
Depending on the distance of the object (typically between 75\,pc and
450\,pc) the spatial resolution required to probe the inner parts of
disks ranges between fractions and a few tens of milli-arcseconds.
Since the angular resolution of astronomical instruments depends
linearly on the wavelength and inversely on the telescope diameter,
observing in the near and mid infrared wavelength domain points toward
telescopes of sizes ranging from ten to several hundreds of meters.
The only technique that allows such spatial resolution is therefore
infrared interferometry.

\citet{2007prpl.conf..539M}, published in \emph{Protostars and
  Planets V}, reviewed the main results obtained with infrared
interferometry in the domain of young stars between 1998 and 2005. I
presented a review talk on this subject at the IAU Symposium 243 on
\emph{Star-disk interaction in young stars}
\citep{2007IAUS..243..123M} as well as R.~Akeson in several schools in
2008 \citep{2008NewAR..52...94A,2008JPhCS.131a2019A}.

The purpose of the present review is to give the latest results in
this field. Section \ref{sect:irinterf} briefly explains the
principles of infrared interferometry and lists the literature on the
observations carried out with this technique. Section
\ref{sect:innerdisks} focuses on the main results obtained on disk
physics (sizes, structures, dust and gas components,...) and
Sect.~\ref{sect:others} presents results on other phenomena
constrained by interferometry (winds, magnetosphereic accretion,
multiple systems,...). In Sect.~\ref{sect:future}, I finish the review
with the type of results that can be expected in the future.

\section{Infrared interferometry}
\label{sect:irinterf}

\subsection{Principle and observations}

Long baseline optical interferometry consists in mixing the light
received from an astronomical source and collected by several
independent telescopes separated from each other by tens or even
hundreds of meters. The light beams are then overlapped and form an
interference pattern if the optical path difference between the
different arms of the interferometer ---taking into account paths from
the source up to the detector--- is smaller than the coherence length
of the incident wave (typically of the order of several microns).
This interference figure is composed of fringes, i.e.\ a succession of
stripes of faint (destructive interferences) and bright (constructive
interferences) intensity. By measuring the contrast of these fringes,
i.e.\ the normalized flux difference between the darkest and brightest
regions, information about the morphology of the observed
astronomical source can be recovered.
\begin{figure}[t]
  \centering
  \parbox{0.4\hsize}{%
    \includegraphics[width=\hsize]{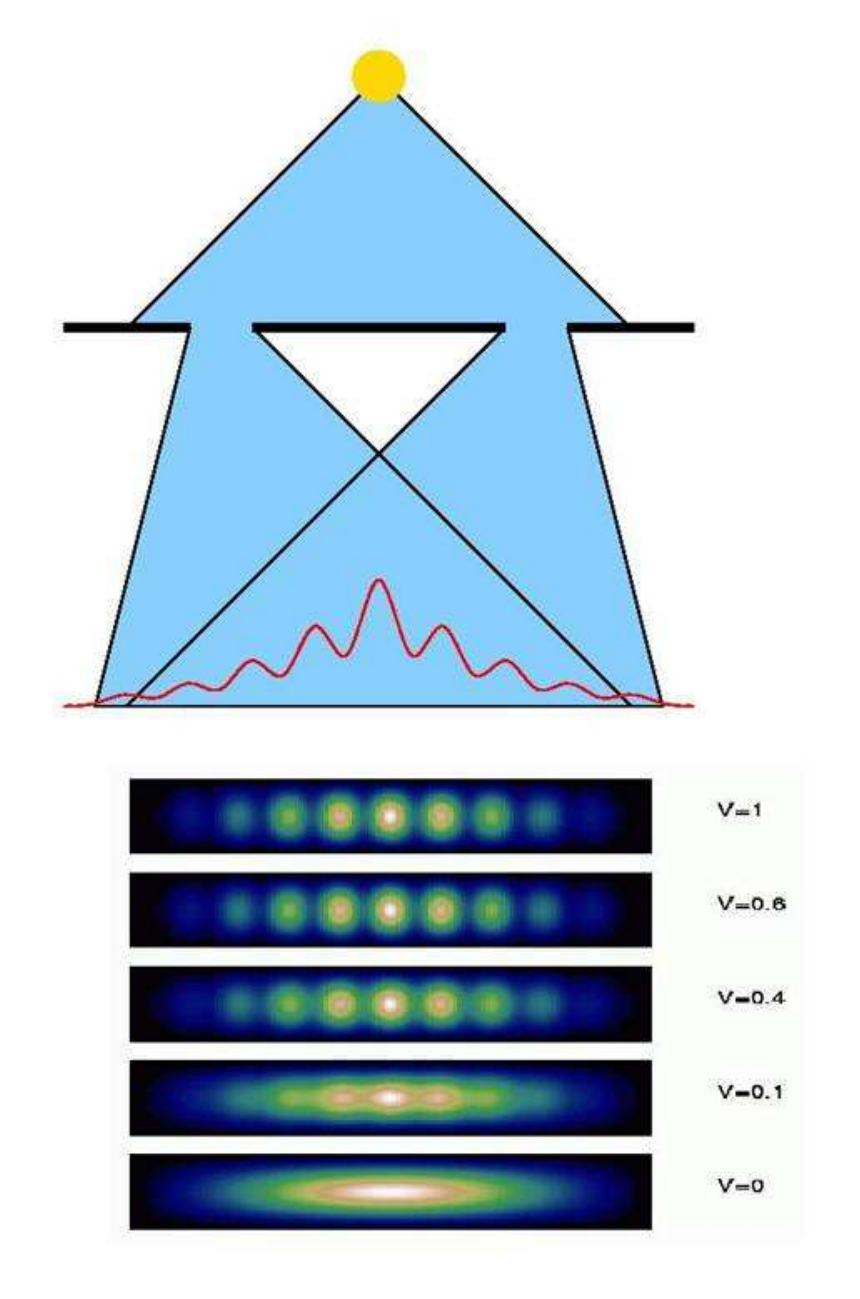}
  }%
  \parbox{0.4\hsize}{%
    \includegraphics[width=0.9\hsize]{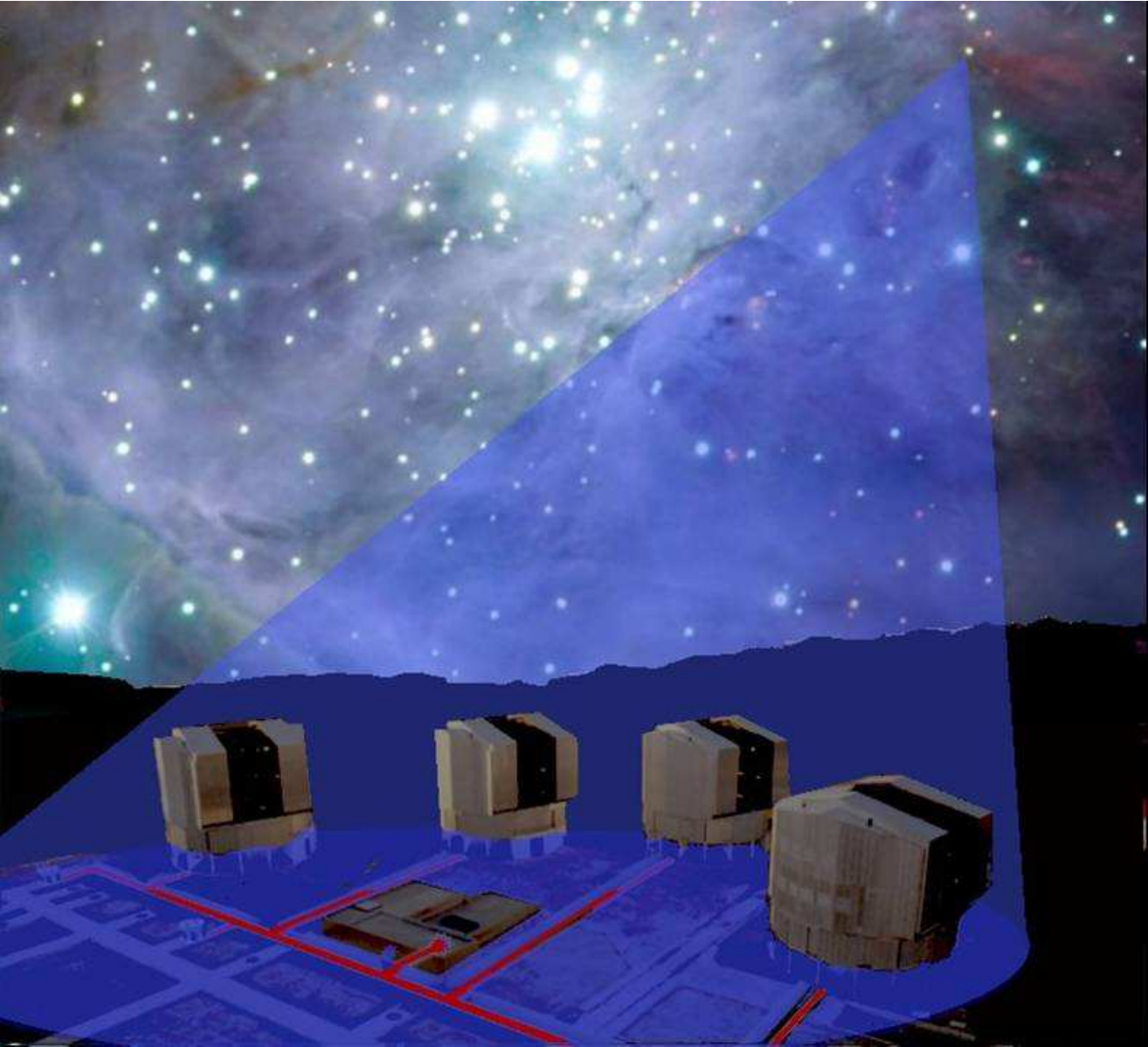}\\
    \includegraphics[width=0.9\hsize]{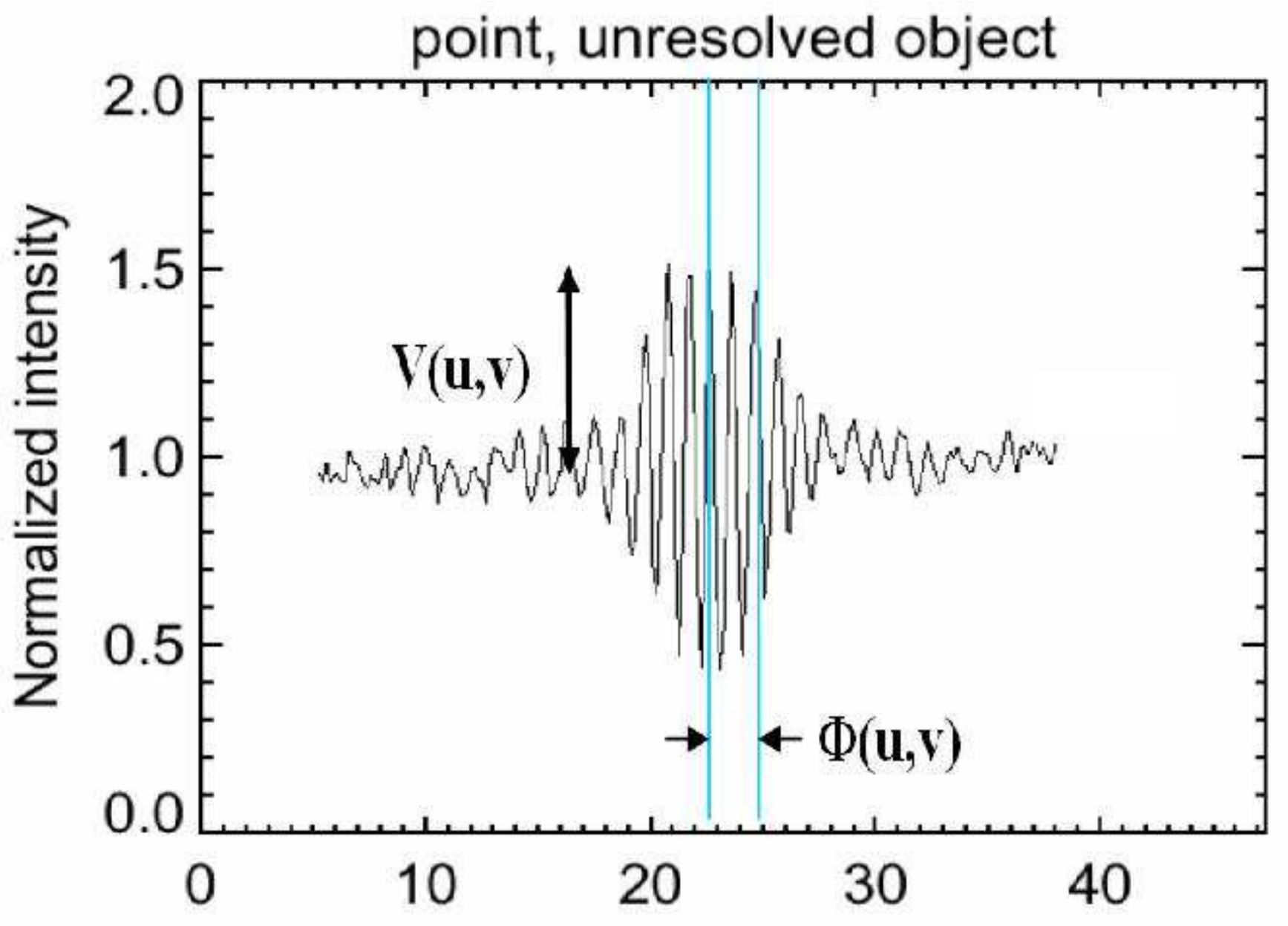}
  }\\
  \caption{Principle of interferometry. Upper panels: Young's slit
    experiment (left) compared to optical interferometry (right): in
    both cases light travels from a source to a plane where the
    incoming wavefront is split. The telescope apertures play the
    same role as the Young's slits. The difference lies in the propagation of
    light after the plane. In the case of optical
    interferometry, the instrument controls the propagation of light
    down to the detectors. At the detector plane, the light beams
    coming from the two apertures are overlapped. Lower panels:
    interference fringes whose contrast changes with the morphology of
    the source. Left panel shows fringes whose contrast varies from 0
    to 1. Right panel displays actual stellar fringes but scanned
    along the optical path. The measure of the complex visibilities
    corresponds to the amplitude of the fringes for the visibility
    amplitude and the position of the fringes in wavelength units for
    the visibility phase.}
  \label{fig:interf}
\end{figure}
Figure \ref{fig:interf} illustrates this principle.

\subsection{Instruments available for inner regions studies}

\begin{figure}[t]
  \centering
  \includegraphics[width=0.8\hsize]{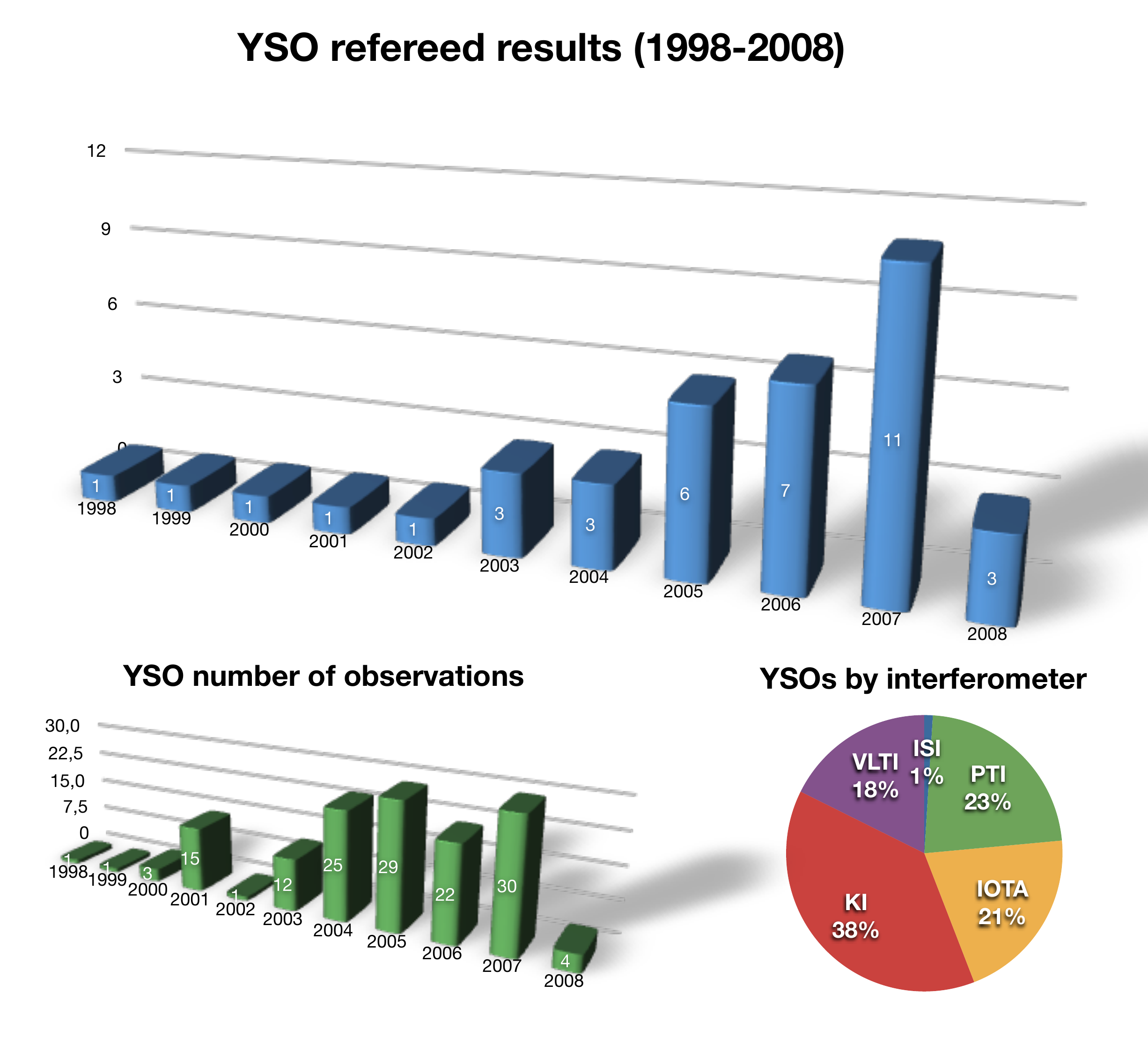}
  \caption{Young stellar objects observed by interferometry and number
    of refereed papers published in the period 1998-2008 (top
    graph). The statistics of the year 2008 is not complete. The
    bottom left graph presents the number of YSO targets observed in
    the same period and the right bottom one gives the distribution by
  interferometer.}
  \label{fig:bib}
\end{figure}

\begin{figure}[t]
  \centering
  \includegraphics[width=0.8\hsize]{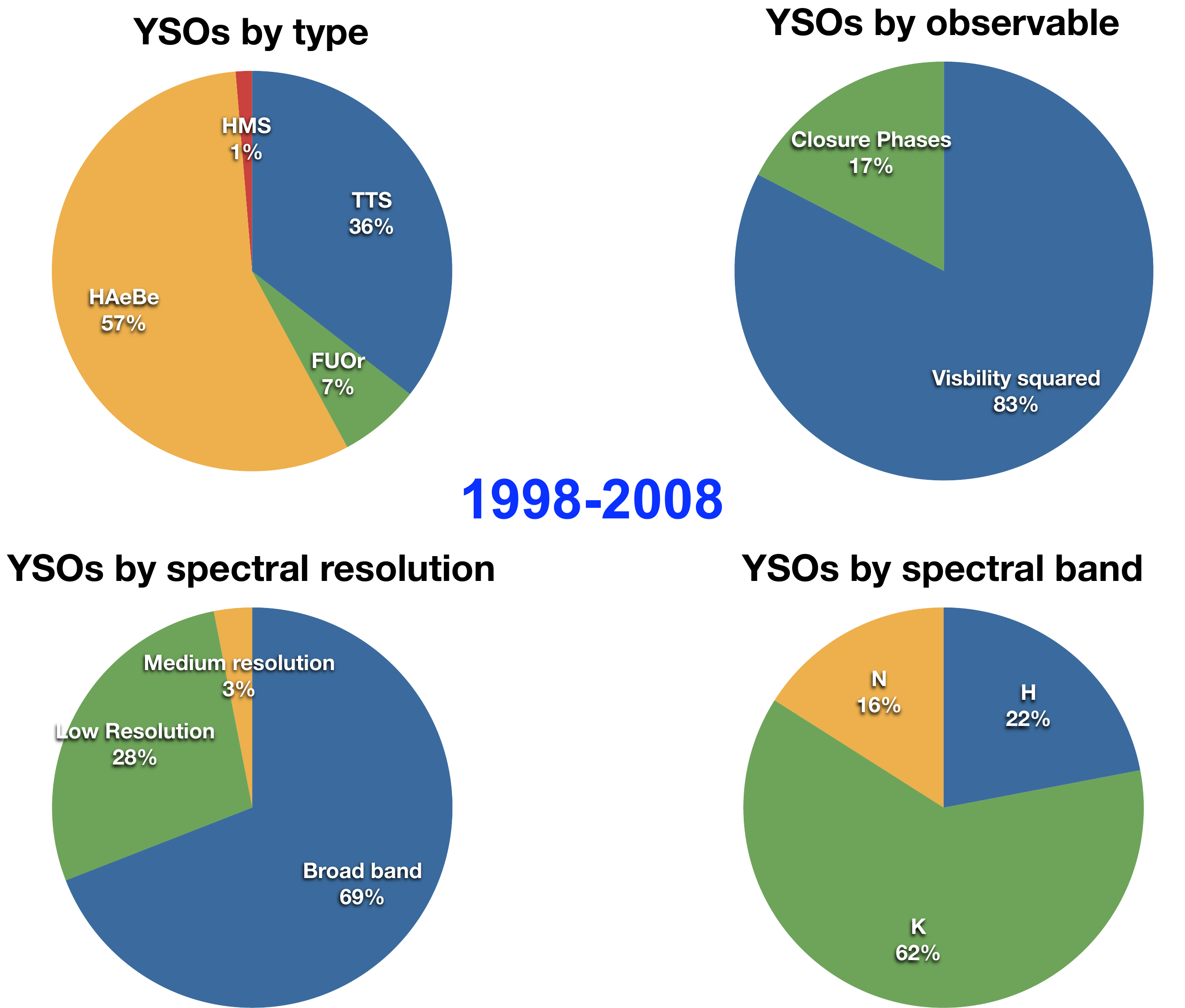}
  \caption{Young stellar objects observed by interferometry in the
    period 1998-2008. Upper left: distribution by YSO type. Upper
    right: distribution by type of observable. Lower left:
    distribution by spectral resolution. Lower right: distribution by
    wavelength of observation. The statistics are the same as the one
    of Fig.~\ref{fig:bib}. }
  \label{fig:various}
\end{figure}

\begin{table}[t]
  \renewcommand{\arraystretch}{1.25} 
  \centering
  \caption{Interferometers involved in YSO science}
  \label{tab:interferometers}
  \medskip
  \begin{tabular}{llllll}
    \hline\hline
    Facility &Instrument &Wavelength &Numbers of &Aperture    &Baseline\\
    &observable           &(microns)  &apertures  &diameter (m)&(m)\\
    \hline\hline
    PTI      &$V^2$      &$H$, $K$   &3          &0.4         &$80-110$\\
    \hline
    IOTA     &$V^2$, CP  &$H$, $K$   &3          &0.4         &$5-38$\\
    \hline
    ISI      &heterodyne &11         &2 (3)      &1.65        &$4-70$\\
    \hline
    KI       &$V^2$, nulling &$K$    &2          &10          &$80$\\
    \hline
    VLTI/AMBER &$V^2$, CP&$1-2.5$    &3 (8)      &8.2/1.8     &$40-130$\\
    &(imaging)  &/spectral   &           &            &/$8-200$\\
    \hline
    VLTI/MIDI&$V^2$ (/CP)&$8-13$     &2 (4)      &8.2/1.8     &$40-130$\\
    &           &/spectral  &           &            &/$8-200$\\
    \hline
    CHARA    &$V^2$, CP  &$1-2.5$    &2/4 (6)    &1           &$50-350$\\
    &(imaging)  &/spectral  &           &            &\\
    \hline
    LBT      &imaging,   &$1-10$     &2          &8.4         &$6-23$\\
    &nulling    &           &           &            &\\
    \hline\hline
    \multicolumn{6}{p{0.85\textwidth}}{\smallskip \footnotesize $V^2$:
      visibility measurement; CP: closure phase.}\\
    \multicolumn{6}{p{0.85\textwidth}}{\smallskip \footnotesize
      Acronyms. PTI: \emph{Palomar Testbed Interferometer}; 
      IOTA: \emph{Infrared and Optical Telescope Array} (closed since
      2006); ISI: \emph{Infrared Spatial Interferometer}; KI: \emph{Keck
        Interferometer}; VLTI: \emph{Very Large Telescope
        Interferometer}; CHARA: \emph{Center for High Angular
        Resolution Array}; LBT: \emph{Large Binocular Telescope} (not
      yet operational).}\\
  \end{tabular}
\end{table}
Interferometric observations of young stellar objects were and are
still performed at six different facilities on seven different
instruments (see Table~\ref{tab:interferometers}). We can classify
these observations into three different categories:
\begin{itemize}
\item \textbf{Small-aperture interferometers}: PTI, IOTA and ISI were
  the first facilities to be operational for YSO observations in the
  late 1990's (see Figs.~\ref{fig:bib} and \ref{fig:various}). They
  have provided mainly the capability of measuring visibility
  amplitudes and lately closure phases. The latest one, CHARA, has
  an aperture diameter of 1\,m. The instruments are mainly
  accessible through team collaboration. IOTA was shut down in 2007
  and PTI in January 2009. 
\item \textbf{Large-aperture interferometers}: KI, VLTI and soon LBT
  are facilities with apertures larger than 8\,m. The instruments are
  widely open to the astronomical community through general calls for
  proposals. Lately, these facilities have significantly increased the
  number of young objects observed.
\item \textbf{Instruments with spectral resolution:} CHARA, MIDI and
  AMBER provide spectral resolution from a few hundred up to 10,000
  whereas other instruments mainly provided broadband observations.
  The spectral resolution allows the various phenomena occurring in the
  environment of young stars to be separated.
\end{itemize}


\subsection{Elements of bibliography}

Figure \ref{fig:bib} displays the number of published results, and
show that it is increasing with time and improved facilities. At the
date of the conference there were 38 refereed articles published in
the field of young stars corresponding to 76 young stellar objects
observed\footnote{See the references by chronological order:
  \citep{1998ApJ...507L.149M, 1999ApJ...513L.131M,
    2000ApJ...543..313A, 2001ApJ...546..358M, 2002ApJ...577..826T,
    2003ApJ...588..360E, 2003ApJ...592L..83C, 2003Ap&SS.286..145W,
    2004A&A...423..537L, 2004ApJ...613.1049E, 2004Natur.432..479V,
    2005A&A...437..627M, 2005ApJ...622..440A, 2005ApJ...623..952E,
    2005ApJ...624..832M, 2005ApJ...635..442B, 2005ApJ...635.1173A,
    2006ApJ...637L.133E, 2006ApJ...641..547M, 2006A&A...458..235P,
    2006A&A...449L..13A, 2006ApJ...645L..77M, 2006ApJ...647..444M,
    2006ApJ...648..472Q, 2007ApJ...657..347E, 2007A&A...464...43M,
    2007A&A...464...55T, 2007A&A...466..649K, 2007A&A...469..587L,
    2007Natur.447..562E, 2007A&A...471..173R, 2007ApJ...669.1072E,
    2007ApJ...670.1214B, 2007ApJ...671L.169D, 2008A&A...479..589L,
    2008ApJ...676..490K, 2008ApJ...677L..51T, 2008A&A...483L..13I,
    2008A&A...485..209A}.}.

Graphs in Fig.~\ref{fig:various} show that the distribution of
observed object is rather well distributed among the various
facilities. Several categories of young stellar systems have been
observed at milli-arcsecond scales mainly in the near-infrared
wavelength domain, but also in the mid-infrared one. They include the
brightest Herbig Ae/Be stars, the fainter T Tauri stars and the few FU
Orionis. Finally most observations were carried out in broad band but
the advent of large aperture interferometers like the VLTI and KI
allow higher spectral resolution to be obtained.

\section{Inner disk physics}
\label{sect:innerdisks}

Most YSO studies are focused on the physics of the inner regions of
disks. They started with the determination of rough sizes of emitting
regions and naturally led to more constraints on the disk
structure. Spectrally resolved mid-infrared observations are able to
identify different types of dust grains. Spectrally resolved
near-infrared observations permit to spatially
discriminate between gas and dust.

\subsection{Sizes of circumstellar structures}

Disks are known to be present around young stars. Some ten years ago,
disks were believed to behave ``normally'' with a radial temperature
distribution following a power-law $T \propto r^{-q}$ with $q$ ranging
between 0.5 and 0.75. The value of $q$ depends on the relative effect
of irradiation from the central star in comparison with heat
dissipation due to accretion.  This model was successful to reproduce
ultraviolet and infrared excesses in spectral energy distributions
(SEDs).  \citet{1995A&AS..113..369M} investigated the potential of
optical long baseline interferometry to study the disks of T Tauri
stars and FU Orionis stars. They found that the structure would be
marginally resolved but observations would be possible with baselines
of the order of 100\,m with a visibility amplitude remaining high.

First observations of the brighter Herbig Ae/Be stars showed that the
observed visibilities were much smaller than expected especially for those 
objects where the accretion plays a little role.
\citet{2002ApJ...579..694M} pointed out that the interferometric sizes
of these objects were much larger than expected from the standard disk
model. They plotted the sizes obtained as function of the stellar
luminosity and found that there was a strong correlation following
a $L^{0.5}$ law over two decades. This behavior is consistent with
the variation of radius of dust sublimation with respect to the
central star luminosity: the dust distribution radius shifts to larger radii for more luminous objects because the temperature is larger than the
sublimation limit ($\sim 1000-1500\,\mbox{K}$). Only the most massive
Herbig Be stars seem to be compliant with the standard accretion disk
model.

In the meantime, in order to account for the near-infrared
characteristics of SEDs and in particular a flux excess around
$\lambda=3\microns$, \citet{2001A&A...371..186N} proposed that disks
around Herbig Ae/Be stars have an optically thin inner cavity and create
a puffed-up inner wall of optically thick dust at the dust sublimation
radius. More realistic models were developed afterward which take
more physical properties into account \citep{2001ApJ...560..957D,
  2004ApJ...617..406M, 2005A&A...438..899I}. However as pointed out by
\citet{2003MNRAS.346.1151V} and recently by
\citet{2008A&A...478..779S} on the specific and actual case of RY~Tau,
the disk models are not the only ones that can reproduce the
measurements: models with a disk halo or envelope can also match the
data.

Observations at KI \citep{2003ApJ...592L..83C, 2005ApJ...623..952E,
  2005ApJ...622..440A} found also large NIR sizes for lower-luminosity
T Tauri stars, in many cases even larger than would be expected from
extrapolation of the HAe relation. It is interpreted by the fact that
the accretion disk contributes significantly to the luminosity emitted
by the central region and therefore this additional luminosity must be
taken into account in the relationship . However in these systems
uncertainties are still large and very few measurements per object have been
obtained. In order to interpret all the T Tauri
measurements, \citet{2005ApJ...622..440A} need to introduce a new
physical component like optically thick gas emission in the inner hole
and extended structure around the objects.

Characteristic dimensions of the emitting regions at $10\,\microns$
were found by \cite{2004A&A...423..537L} to be ranging from 1 AU to 10
AU. The sizes of their sample stars correlate with the slope of the
$10-25\,\microns$ infrared spectrum: the reddest objects are the
largest ones.  Such a correlation is consistent with a different
geometry in terms of flaring or flat (self-shadowed) disks for sources
with strong or moderate mid-infrared excess, respectively,
demonstrating the power of interferometry not only to probe
characteristic sizes of disks but also to derive information on the
vertical disk structure.

\subsection{Constraints on disk structure}
\label{sect:diskstructure}

Theoreticians start discussing slightly different scenarios of the
inner regions around young stars. For example, the inner
puffed-up wall is modeled with a curved shape by
\citet{2005A&A...438..899I} due to the very large vertical density
gradient and the dependence of grain evaporation temperature on gas
density as expected when a constant evaporation temperature is
assumed. \citet{2007ApJ...661..374T} proposed that the
geometry of the rim depends on the composition and spatial
distribution of dust due to grain growth and settling.

\citet{2007ApJ...658..462V} presented a model-independent method of
comparison of NIR visibility data of YSOs. The method based on scaling
the measured baseline with the YSO distance and luminosity removes the
dependence of visibility on these two variables. They found that low
luminosity Herbig Ae stars are best explained by the uniform
brightness ring and the halo model, T Tauri stars with the halo model,
and high luminosity Herbig Be stars with the accretion disk model, but
they admit that the validity of each model is not well established.

At the moment, only two objects have been thoroughly studied:
FU Orionis \citep{1998ApJ...507L.149M, 2005A&A...437..627M,
  2006ApJ...648..472Q} and MWC\,147 \citep{2008ApJ...676..490K}.
FU\,Ori has been observed on 42 nights over a period of 6 years from
1998 to 2003 with 287 independent measurements of the fringe
visibility at 6 different baselines ranging from 20 to 110 m in
length, in the $H$ and $K$ bands. The data not only resolves FU Ori at
the AU scale, but also allows the accretion disk scenario to be
tested. The most probable interpretation is that FU Ori hosts an
active accretion disk whose temperature law is consistent with the
standard model. In the mid infrared, \citet{2006ApJ...648..472Q}
resolved structures that are also best explained with an optically
thick accretion disk. A simple accretion disk model fits the observed
SED and visibilities reasonably well and does not require the presence
of any additional structure such as a dusty envelope. However
\citet{2008ApJ...684.1281Z} revisited this issue using detailed
radiative transfer calculations to model the recent, high
signal-to-noise ratio data obtained on FU\,Ori from the IRS instrument
on the Spitzer Space Telescope. They found that a physically plausible
flared disk irradiated by the central accretion disk matches the
observations. Their accretion disk model with outer disk irradiation
by the inner disk reproduces the spectral energy distribution and
their model is consistent with near-infrared interferometry, but there
remains some inconsistencies with mid-infrared interferometric
results. This is why one should remain careful with results coming
from surveys having only few measurements per object.

\citet{2006ApJ...641..547M} obtained $K$-band observations of three
other FU Orionis objects, V1057 Cyg, V151 Cyg, and Z CMa-SE and found
that all three objects appear significantly more resolved than
expected from simple models of accretion disks tuned to fit the SEDs.
They believe that emission at the scale of tens of AU in the
interferometer field of view is responsible for the low visibilities,
originating in scattering by large envelopes surrounding these
objects. \citet{2008A&A...479..589L} have measured
again interferometric visibilities of Z CMa with VLTI/AMBER.

\begin{figure}[p]
  \centering 
  \includegraphics[width=0.8\hsize]{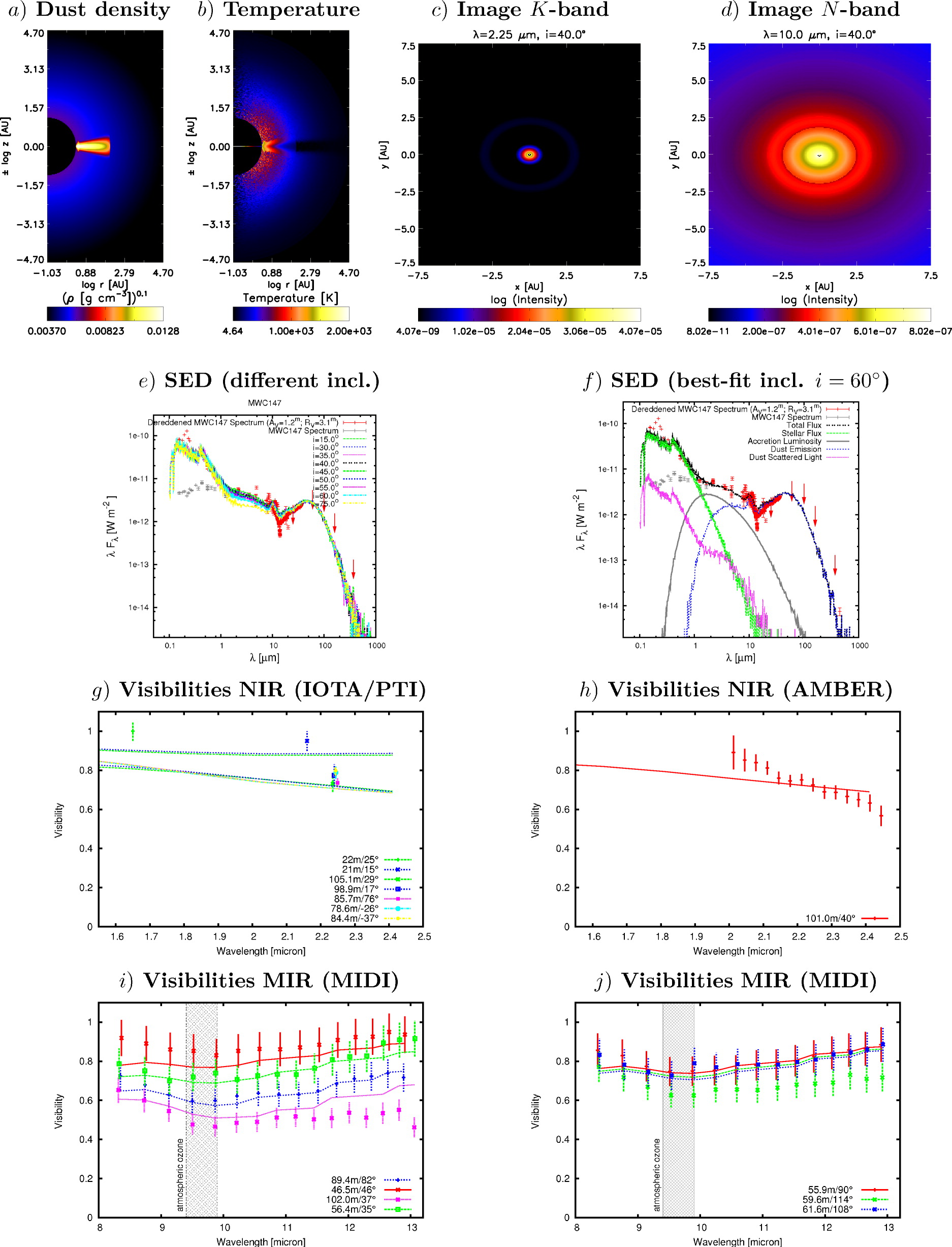}
  \caption{Radiative transfer model constrained by interferometric
    measurements on MWC\,147 by \citet{2008ApJ...676..490K}. The model
    computed for MWC 147 assumes a spherical shell geometry. Panels
    (a) and (b) show the dust density and the temperature
    distribution. Panels (c) and (d) show the ray-traced images for
    two representative NIR ($2.25\,\microns$) and MIR
    ($10.0\,\microns$) wavelengths. Panel (e) shows the SED for
    various inclination angles, whereas panel (f) gives the SED for
    the best-fit inclination angle and separates the flux which
    originates in stellar photospheric emission, thermal emission,
    dust irradiation, and accretion luminosity. Finally, panels (g) to
    (j) depict the NIR and MIR visibilities computed from their
    radiative transfer models.}
  \label{fig:kraus08}
\end{figure}

\citet{2008ApJ...676..490K} have shown that they are able to derive the
temperature radial distribution of the disk around MWC\,147 from the
interferometric measurements using the spectral variation of the
visibilities in low resolution with AMBER and MIDI (see Fig.~\ref{fig:kraus08}). A similar work has been attempted at
PTI with larger uncertainties \citep{2007ApJ...657..347E}.   

On the pure theoretical side, very few physical models achieved to fit
interferometric data simultaneously with SEDs. Using a two-layer
accretion disk model, \citet{2003A&A...400..185L} found satisfactory
fits for SU Aur, in solutions that are characterized by the midplane
temperature being dominated by accretion, while the emerging flux is
dominated by reprocessed stellar photons.  Since the midplane
temperature drives the vertical structure of the disk, there is a
direct impact on the measured visibilities that are not necessarily
taken into account by other models. In the MIR range,
\citet{2008A&A...478..779S} was able to fit both the SED and the
visibilities of RY~Tau as mentioned earlier with 2 different models
although several other models have been dismissed with MIDI data. The
region where the measurements would allow us to choose between the two
remaining models is located inward, and can be observed at shorter
wavelengths.

\subsection{Dust mineralogy}

The mid-infrared wavelength region contains strong resonances of
abundant dust species, both oxygen-rich (amorphous or crystalline
silicates) and carbon-rich (polycyclic aromatic hydrocarbons,
or PAHs). Therefore, spectroscopy of optically thick
protoplanetary disks offers a diagnostic of the chemical
composition and grain size of dust in disk atmospheres. 

\citet{2004Natur.432..479V} spatially resolved three protoplanetary
disks surrounding Herbig Ae/Be stars across the $N$ band. The
correlated spectra measured by MIDI at the VLTI correspond to disk
regions ranging from 1 to 2\,AUs. By combining these measurements with
unresolved spectra, the spectrum corresponding to outer disk regions
at 220 AU can also be derived. These observations have revealed that
the dust in these regions was highly crystallized (40 to 100\%), more
than any other dust observed in young stars until now. The spectral
shape of the inner-disk spectra shows surprising similarity with Solar
System comets. Their observations imply that silicates crystallize
before terrestrial planets are formed, consistent with the
composition of meteorites in the Solar System. Similar measurements
were also carried out by \citet{2007A&A...471..173R} on the T Tauri
system, TW Hya.  According to the correlated flux measured with MIDI,
most of the crystalline material is located in the inner, unresolved part of
the disk, about 1\,AU in radius.

\subsection{Gas/dust connection}

\citet{2008poii.conf..187G} observed the young stellar system 51 Oph
confirming the interpretation of \citet{2005A&A...430L..61T} and more
recently \citet{2007ApJ...660..461B} of a disk seen nearly edge-on:
the radial distribution of excitation temperatures for the vibrational
levels of CO overtone ($\Delta v=2$) emission from hot gas is
consistent with the gas being in radiative thermal equilibrium except
at the inner edge, where low vibrational bands have higher excitation
temperatures. \citet{2008A&A...489.1151T} confirmed the high
inclination of the disk but also detect the CO bandheads allowing the
dust responsible for the continuum to be separated from the gas. As a
matter of fact, the visibilities in the CO bands is lower than the
ones measured in the continuum implying that the region responsible
for this gas emission is smaller than the region responsible for the
dust emission.
\begin{figure}
  \centering
  \includegraphics[width=0.5\hsize]{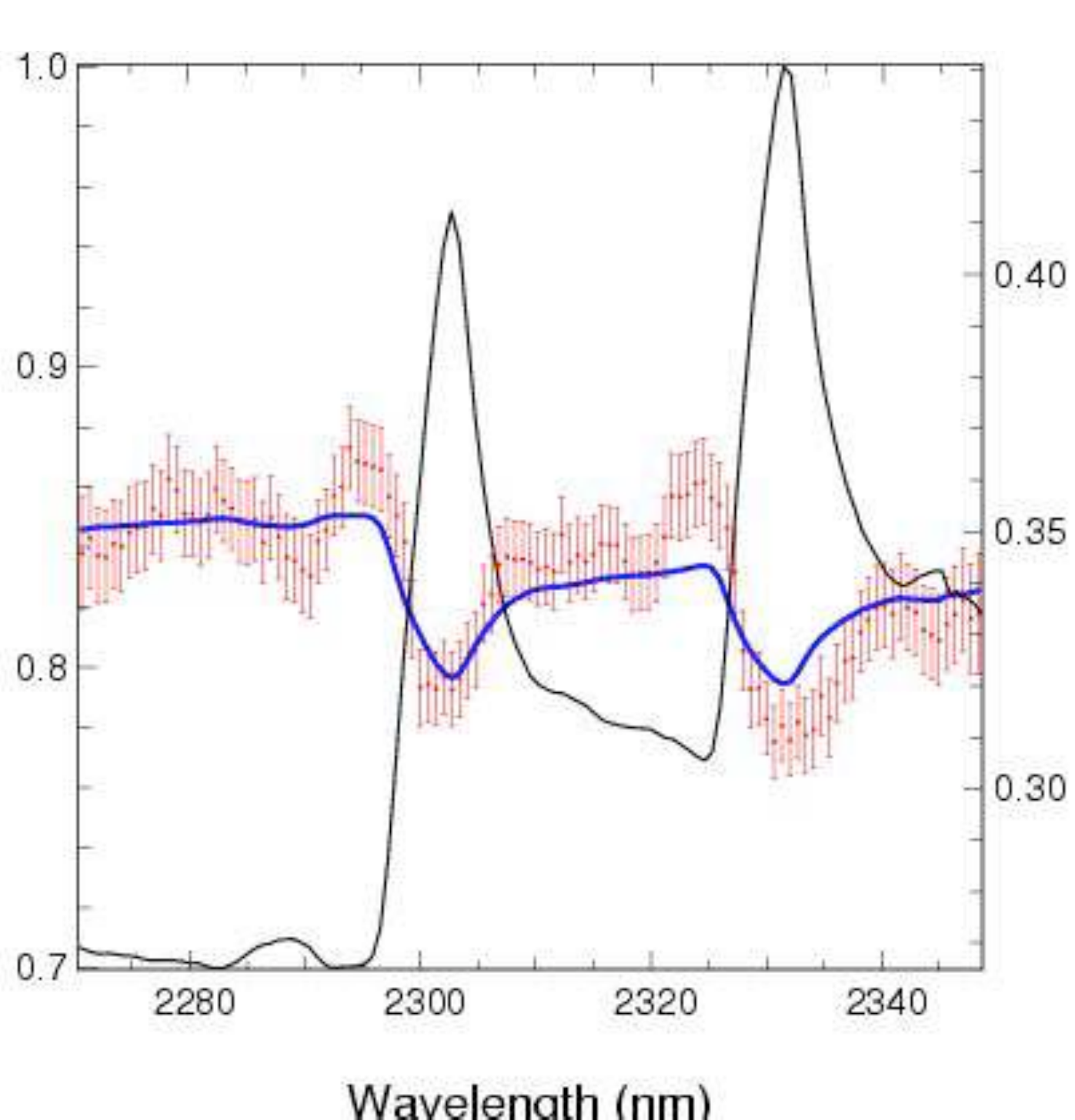}
  \caption{Spectrally dispersed visibility amplitudes of 51 Oph in the
    CO bandhead spectral region. Overimposed is the spectrum as
    measured by VLTI/AMBER (black line). The blue curve corresponds to
    the addition of a simple uniform disk model for the excess emission
    in the line with a typical diameter of 0.2\,AU. From Tatulli et
    al.\ (priv.\ comm.)}
  \label{fig:51oph}
\end{figure}
Figure \ref{fig:51oph} illustrates this result, and shows that the
combination of very high spatial information with spectral resolution
opens brand new perspectives in the studies of the inner disk
properties by discriminating between components.

\section{Other AU-scale phenomena}
\label{sect:others}

Several other physical phenomena have been investigated in the
innermost region of disks: wind, magnetosphere and close companions.

\subsection{Outflows and winds}

The power of spectrally resolved interferometric measurements provides
detailed wavelength dependence of inner disk continuum emission (see
end of Sect.~\ref{sect:diskstructure}). These new capabilities enable
also detailed studies of hot winds and outflows, and therefore the
physical conditions and kinematics of the gaseous components in which
emission and absorption lines arise like Br$\gamma$ and H$_2$ ones.
With VLTI/AMBER, \citet{2007A&A...464...43M} spatially resolved the
luminous Herbig Be object MWC 297, measuring visibility amplitudes as
a function of wavelength at intermediate spectral resolution R = 1500
across a $2.0-2.2\,\microns$ band, and in particular the Br$\gamma$
emission line. The interferometer visibilities in the Br$\gamma$ line
are about 30\% lower than those of the nearby continuum, showing that
the Br$\gamma$ emitting region is significantly larger than the NIR
continuum region.  Known to be an outflow source, a preliminary model
has been constructed in which a gas envelope, responsible for the
Br$\gamma$ emission, surrounds an optically thick circumstellar disk.
The characteristic size of the line-emitting region being 40\% larger
than that of the NIR disk.  This model is successful at reproducing
the VLTI/AMBER measurements as well as previous continuum
interferometric measurements at shorter and longer baselines
\citep{2001ApJ...546..358M, 2004ApJ...613.1049E}, the SED, and the
shapes of the H$\alpha$, H$\beta$, and Br$\gamma$ emission lines.  The
precise nature of the MWC 297 wind, however, remains unclear; the
limited amount of data obtained in these first observations cannot,
for example, discriminate between a stellar or disk origin for the
wind, or between competing models of disk winds
\citep[e.g.][]{2007IAUS..243..249S,2007IAUS..243..307F}.

\subsection{Magnetosphere}

\begin{figure}[t]
  \centering
  \includegraphics[width=0.55\hsize]{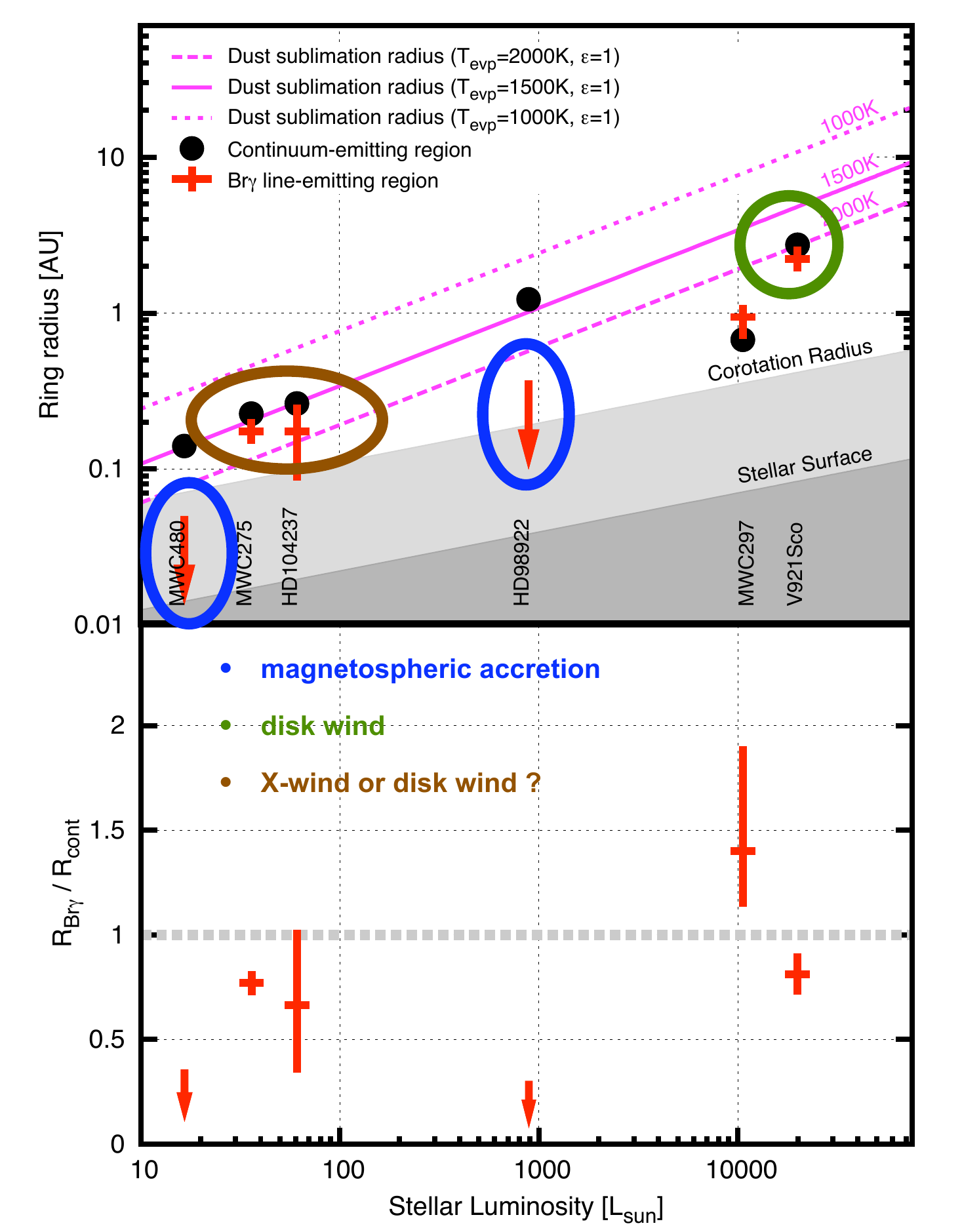}
  \caption{A systematic study of the origin of the Br$_\gamma$
    emission in Herbig Ae/Be stars from
    \citet{2007ApJ...670.1214B}. Top: the fitted ring radii for the
    continuum (black points) and Br$_\gamma$ line (red points) plotted
    as a function of stellar luminosity. The spatial extension of the
    stellar surface is represented in grey and the dust sublimation
    radius corresponding to the dust sublimation temperatures of 2000,
    1500, and 1000 K.  Bottom: the ratio of normalized size of the
    Br$_\gamma$ region by the size of the continuum-emitting
    region. Circles in blue emphasize the system for which a
    magnetospheric accretion scenario is a compatible model, in green
    the ones with probable disk winds and in brown the ones with X
    wind or disk wind.}
  \label{fig:surveyhaebes}
\end{figure}

The origin of the hydrogen line emission in Herbig Ae/Be stars is
still unclear. The lines may originate either in the gas which
accretes onto the star from the disk, as in magnetospheric accretion
models \citep{1994ApJ...426..669H}, or in winds and jets, driven by
the interaction of the accreting disk with a stellar
\citep{1994ApJ...429..781S} or disk \citep{2000A&A...353.1115C}
magnetic field. For all models, emission in the hydrogen lines is
predicted to occur over very small spatial scales, a few AUs at most.
To understand the physical processes that happen at these scales, one
needs to combine very high spatial resolution with enough spectral
resolution to resolve the line profile. 

On the one hand, \citet{2007A&A...464...55T} performed interferometric
observations of the Herbig Ae star HD~104237, obtained with the
VLTI/AMBER instrument with $R = 1500$ high spectral resolution. The
observed visibility was identical in the Br$\gamma$ line and in the
continuum, even though the line represents 35\% of the continuum flux.
This immediately implies that the line and continuum emission regions
have the same apparent size.  Using simple models to describe the
Br$\gamma$ emission, they showed that the line emission is unlikely to
originate in either magnetospheric accreting columns of gas or in the
gaseous disk but more likely in a compact outflowing disk wind
launched in the vicinity of the rim, about 0.5\,AU from the star. The
main part of the Br$\gamma$ emission in HD~104237 is unlikely to
originate in magnetospheric accreting matter.

On the other hand, \citet{2007Natur.447..562E} measured an increase of
the Br$\gamma$ visibility in MWC 480 implying that the region of
emission of the hydrogen line is very compact, less than 0.1\,mas in
radius which could be interpreted as emission originating in the
magnetosphere of the system.  

At the present time, given the limited number of samples, it is
difficult to derive a general tendency but it seems that all possible
scenari can be found like \citet{2008A&A...489.1157K} show in a
mini-survey of five Herbig Ae/Be stars.

\subsection{Binaries and multiple systems}

\citet{2005ApJ...635..442B} performed the first direct measurement of
pre-main sequence stellar masses using interferometry, for the
double-lined system HD 98800-B. These authors established a
preliminary orbit that allowed determination of the (subsolar) masses
of the individual components with 8\% accuracy.  Comparison with
stellar models indicates the need for subsolar abundances for both
components, although stringent tests of competing models will only
become possible when more observations improve the orbital phase
coverage and thus the accuracy of the stellar masses
derived. \citet{2007ApJ...670.1214B} published another determination
of dynamical masses for the pre-mains sequence system V773\,Tau\,A.

Another example, based on a low-level oscillation in the visibility
amplitude signature in the PTI data of FU Ori,
\citet{2005A&A...437..627M} claimed the detection of an off-centered
spot embedded in the disk that could be physically interpreted as a
young stellar or protoplanetary companion located at $\sim
10\,\mbox{AUs}$, and could possibly be at the origin of the FU Ori
outburst itself. Using another technique, \citet{2006ApJ...645L..77M}
reported on the detection of localized off-center emission at 1-4 AU
in the circumstellar environment of AB Aurigae. They used
closure-phase measurements in the near-infrared. When probing sub-AU
scales, all closure phases are close to zero degrees, as expected
given the previously determined size of the AB Aurigae inner-dust
disk. However, a clear closure-phase signal of
$-3.5^{\circ}\pm0.5^{\circ}$ is detected on one triangle containing
relatively short baselines, requiring a high degree of asymmetry from
emission at larger AU scales in the disk. They interpret such detected
asymmetric near-infrared emission as a result of localized viscous
heating due to a gravitational instability in the AB Aurigae disk, or
to the presence of a close stellar companion or accreting substellar
object.

\section{Future prospects and conclusion}
\label{sect:future}

As emphasized in this review, more interferometric data is required
with better accuracy and also wider coverage of the baselines in order
to constrain better the models that have been proposed. Like for radio
astronomy, these supplementary data will allow image reconstruction
without any prior knowledge of the observed structure. Several
facilities are already ready to obtain interferometric images although
with few pixels across the field: MIRC at CHARA and AMBER at the VLTI
in the near-infrared. However at the moment MIRC is limited in
sensitivity and AMBER in number of telescopes (3) which makes it
difficult to routinely image. In the mid-infrared the
MATISSE instrument has been proposed to ESO to provide imaging with 4
telecopes at the VLTI. VSI is also a proposed second generation VLTI instrument
which will be able to combine from 4 to 8 beams at the same time
so that imaging becomes easier. The LBT will also provide imaging capability.

All these instruments provide spectral resolution that make them
indeed spectro-imagers. Therefore in the future, one should be able to
obtain a wealth of information from the innermost regions of disks
around young stars. However in the meantime, observations are already
mature enough to allow detailed modeling of the phenomena occuring in
these inner regions of young stellar objects.

\emph{Acknowledgements. The author is grateful to Willem-Jan de Wit for a careful reading of the manuscript and his fruitful comments.}

\bibliographystyle{elsarticle-harv}
\bibliography{ysos-vlti-malbet}





\end{document}